\documentclass{article}
\usepackage[utf8]{inputenc}
\pdfoutput=1
\usepackage{amsmath}
\usepackage{enumerate}
\usepackage{amssymb}
\usepackage{float}
\usepackage{graphicx}
\usepackage{multirow}
\usepackage{xcolor}
\usepackage{bm}
\usepackage[normalem]{ulem}
\usepackage[autostyle]{csquotes} 

\usepackage{booktabs}
\usepackage{color}
\usepackage{jcappub}
\usepackage{orcidlink}

\newcommand{\EF}[2]{#1 }

\title{
Assembly bias and local Primordial non-Gaussianity from DESI DR1 Quasars}
\author[1]{E.~Fondi\orcidlink{0000-0001-7801-1859 },}
\author[1,2]{L.~Verde\orcidlink{0000-0003-2601-8770 },}
\author[3]{E.~Chaussidon\orcidlink{0000-0001-8996-4874 },}
\author[3]{J.~Aguilar,}
\author[4]{S.~Ahlen\orcidlink{0000-0001-6098-7247 },}
\author[5]{S.~BenZvi\orcidlink{0000-0001-5537-4710 },}
\author[6,7]{D.~Bianchi\orcidlink{0000-0001-9712-0006 },}
\author[8]{D.~Brooks,}
\author[3]{T.~Claybaugh,}
\author[3]{A.~Cuceu\orcidlink{0000-0002-2169-0595 },}
\author[9]{A.~de la Macorra\orcidlink{0000-0002-1769-1640 },}
\author[8]{P.~Doel,}
\author[3,10]{S.~Ferraro\orcidlink{0000-0003-4992-7854 },}
\author[11,12]{J.~E.~Forero-Romero\orcidlink{0000-0002-2890-3725 },}
\author[13,14,15]{E.~Gazta\~naga\orcidlink{0000-0001-9632-0815 },}
\author[16]{S.~Gontcho A Gontcho\orcidlink{0000-0003-3142-233X },}
\author[17]{G.~Gutierrez,}
\author[18,19]{H.~K.~Herrera-Alcantar\orcidlink{0000-0002-9136-9609 },}
\author[20,21]{D.~Huterer\orcidlink{0000-0001-6558-0112 },}
\author[22]{M.~Ishak\orcidlink{0000-0002-6024-466X },}
\author[23]{R.~Joyce\orcidlink{0000-0003-0201-5241 },}
\author[3]{A.~Kremin\orcidlink{0000-0001-6356-7424 },}
\author[8]{O.~Lahav,}
\author[24]{C.~Lamman\orcidlink{0000-0002-6731-9329 },}
\author[3]{M.~Landriau\orcidlink{0000-0003-1838-8528 },}
\author[25]{L.~Le~Guillou\orcidlink{0000-0001-7178-8868 },}
\author[26,27]{M.~Manera\orcidlink{0000-0003-4962-8934 },}
\author[28,29,24]{P.~Martini\orcidlink{0000-0002-4279-4182 },}
\author[23]{A.~Meisner\orcidlink{0000-0002-1125-7384 },}
\author[2,27]{R.~Miquel,}
\author[14]{S.~Nadathur\orcidlink{0000-0001-9070-3102 },}
\author[19,3]{N.~Palanque-Delabrouille\orcidlink{0000-0003-3188-784X },}
\author[30,31,32]{W.~J.~Percival\orcidlink{0000-0002-0644-5727 },}
\author[33]{F.~Prada\orcidlink{0000-0001-7145-8674 },}
\author[34]{I.~P\'erez-R\`afols\orcidlink{0000-0001-6979-0125 },}
\author[35]{G.~Rossi,}
\author[36,37,38]{L.~Samushia\orcidlink{0000-0002-1609-5687 },}
\author[39]{E.~Sanchez\orcidlink{0000-0002-9646-8198 },}
\author[3]{D.~Schlegel,}
\author[23]{D.~Sprayberry,}
\author[21]{G.~Tarl\'{e}\orcidlink{0000-0003-1704-0781 },}
\author[23]{B.~A.~Weaver,}
\author[40]{H.~Zou\orcidlink{0000-0002-6684-3997}}

\affiliation{Affiliations are in Appendix~\ref{app:affiliations}}

\emailAdd{emanuelefondi@icc.ub.edu, liciaverde@icc.ub.edu}

\abstract{
The analysis of the large-scale clustering of quasars (QSO) observed by the Dark Energy Spectroscopic Instrument (DESI) represents a promising avenue for constraining local Primordial non-Gaussianity (PNG), parameterized by $f_{\rm NL}$. The signal to be constrained is the scale-dependent bias induced in the 2-point clustering of the considered tracer sample. The resulting constraints on $f_{\rm NL}$, however, are fully degenerate with the local PNG bias parameter $b_{\phi}$,
dependent on the assembly bias 
parameter $p$. Using IllustrisTNG hydrodynamical simulations, we select a QSO sample reflecting the selection criteria and properties of DESI QSOs, and provide a robust prior for $p$, and thus for $b_{\phi}$, building on the findings of Fondi et al. 2024.
We find a distribution with mean $\bar{p}\simeq1.4$ with weak redshift dependence, stable to selection noise and consistent with the expected recent merger history typical of quasar-hosting halos. By comparing with the CAMELS simulations we demonstrate that this prior is robust to astrophysical assumptions and cosmic variance. Finally, applying this prior to the DESI DR1 dataset, we derive updated constraints on local PNG, obtaining $f_{\rm NL}=-3.3\pm9.2$.
}

\begin{document}
\maketitle
\section{Introduction}
The inflationary paradigm represents the standard framework for explaining the early evolution of the Universe and the origin of cosmic structures. However, the nature of inflation is still an open question and different scenarios can be constrained by probing the statistical properties of primordial fluctuations. While single-field slow-roll inflation predicts nearly Gaussian primordial fluctuations \cite{maldacena_non-gaussian_2003}, more complex scenarios can generate measurable amounts of primordial non-Gaussianity (PNG). Among various types of PNG, the \textit{local-type} \cite{gangui_three--point_1994, bernardeau_non-gaussianity_2002} emerges naturally in multifield inflationary models and is usually parameterized as:
\begin{equation}
\Phi(\mathbf{x}) = \phi(\mathbf{x}) + f_{\rm NL}\left(\phi^2(\mathbf{x}) - \langle\phi^2(\mathbf{x})\rangle\right),
\end{equation}
where $\Phi$ is the primordial gravitational potential, $\phi$ is a Gaussian random field, and $f_{\rm NL}$ quantifies the amplitude of non-Gaussianity \cite{verde_large-scale_2000,komatsu_acoustic_2001}. 
Cosmic Microwave Background (CMB) anisotropies measurements from Planck \cite{planck_collaboration_planck_2019} currently provide the tightest constraints on local PNG: $f_{\rm NL}=-0.9 \pm 5.1$. Nevertheless, the information accessible through the two-dimensional CMB map, especially on large angular scales, is approaching the
cosmic variance limit. On the other hand, the statistical power of the three-dimensional distribution of tracers probed by galaxy surveys is increasing, having the potential to drastically improve constraints on PNG. Crucially, local PNG induces correlations between small-scale density fluctuations and long-wavelength potential modes, resulting in a distinctive scale-dependent bias in the clustering of dark matter halos and galaxies on large scales \cite{dalal_imprints_2008,matarrese_effect_2008,slosar_constraints_2008,afshordi_primordial_2008,mcdonald_primordial_2008}. This feature represents a clear signature that galaxy and quasar surveys can exploit to constrain multifield inflation \cite{giannantonio_constraining_2012,alvarez_testing_2014,de_putter_designing_2017,karagiannis_constraining_2018,ferraro_snowmass2021_2022,achucarro_inflation_2022}.

In practice, the large-scale power spectrum is sensitive to the product $f_{\rm NL}\,b_\phi$, where $b_{\phi}$ measures the response of tracer number density fluctuations to large-scale primordial potential fluctuations. Therefore, extracting robust constraints on $f_{\rm NL}$ critically depends on the knowledge of $b_{\phi}$. Its conventional prediction, widely adopted in the literature, can be expressed as \cite{dalal_imprints_2008,matarrese_effect_2008, slosar_constraints_2008}:
\begin{equation}\label{eq:universality_relation}
b_{\phi} = 2 \delta_c (b_1 - p),
\end{equation}
where $b_1$ is the linear bias and $\delta_c \approx 1.686$ is the spherical collapse threshold. Typically, $p$ is fixed to $p=1$, in which case Eq.~\eqref{eq:universality_relation} is known as \textit{universality relation}. This relation is valid only for a \textit{fair} sample of halos of mass $M$, the probability distribution of which is given by a universal mass function \cite{bond_excursion_1991,press_formation_1974,sheth_large_1999}. However, departures from universality are expected due to \textit{assembly bias}, whenever the selection of tracers depends on quantities beyond dark matter halo mass \cite{Barreira2020,Barreira2022,Lazeyras2023,Sullivan2023,Barreira2023,Adame2024}. For example, for recently merged halos, early work found a value of $p=1.6$ \cite{slosar_constraints_2008}.

This result was obtained as a particular case of the extended Press-Schechter (ePS) formalism. This framework, in which the conditional mass function is used to incorporate merger history information beyond mass alone \cite{bond_excursion_1991, lacey_merger_1993,lacey_merger_1994}, was then generalized to model $p$ as a function of arbitrary halo formation redshift and validated against $N$-body simulations \cite{reid_non-gaussian_2010, Fondi}.

Quasi-stellar objects (QSO, or quasars), which are assumed to form in recently merged halos, constitute a biased selection relative to a mass-selected sample. Consequently, the choice $p=1.6$ was adopted to analyze the clustering of quasars from the final data release of the extended Baryon Oscillation Spectroscopic Survey (eBOSS), yielding $-23<f_{\rm NL}<21$ at 68\% confidence level~\cite{cagliari2024}. More recently, the first-year data release of the Dark Energy Spectroscopic Instrument (DESI) \cite{Snowmass2013.Levi,DESI2016a.Science,DESI2016b.Instr,DESI2022.KP1.Instr,FocalPlane.Silber.2023,Corrector.Miller.2023,Spectro.Pipeline.Guy.2023,SurveyOps.Schlafly.2023,LRG.TS.Zhou.2023,DESI2023a.KP1.SV,DESI2023b.KP1.EDR,2024arXiv240403000D,2024arXiv240403001D,adame2024desi,2024arXiv240403002D,FiberSystem.Poppett.2024} has provided new measurements using its quasar sample, improving both the statistical power and control over systematics~\cite{desi_fnl2024}.

DESI analysis adopted two values for $p$: a universality choice, $p=1$, and a recent-merger choice, $p=1.6$, implying two different values of $b_{\phi}$ at fixed $b_1$. Since the scale-dependent bias signature constrains only the product $f_{\rm NL}\,b_{\phi}$, $f_{\rm NL}$ is completely degenerate with $b_{\phi}$; consequently, DESI reports two separate constraints on $f_{\rm NL}$, one for each value of $p$~\cite{desi_fnl2024}.

Motivated by these considerations, in this paper we use the state-of-the-art IllustrisTNG \cite{nelson2018first,nelson_illustristng_2021} hydrodynamical simulations and the CAMELS simulations \cite{villaescusa-navarro_camels_2021} to construct a QSO sample reflecting the selection properties of the DESI survey. Leveraging the ePS framework and the methodology of Ref.~\cite{Fondi}, we calibrate a robust and physically motivated prior $\Pi(p)$ for the parameter $p$, thus breaking the degeneracy between $f_{\rm NL}$ and $b_{\phi}$. Using $\Pi(p)$, we then derive a single updated constraint on local PNG from the DESI DR1 QSO dataset.

The paper is structured as follows: Section 2 presents the theoretical background, introducing the ePS prescription that will be used to model $p$ from the formation redshift. In Section 3, we build a DESI-like QSO sample in IllustrisTNG and study its merger history, providing a distribution for $p$ and testing its robustness to selection effects. Section 4 validates the prior with CAMELS, quantifying sensitivity to cosmic variance and quasar physics. In Section 5, we apply the simulation-based prior to DESI DR1 and report the updated constraint on $f_{\rm NL}$. Finally, conclusions are outlined in Section 6. Appendix \ref{app:redshift_dependence} documents the redshift dependence of the prior.

\section{Theoretical background}\label{sec:background}
The large-scale modulation induced by local PNG can be described analytically in the peak-background split picture \cite{slosar_constraints_2008}. In this approach, the long mode acts a background rescaling of the local amplitude of small-scale clustering (quantified by $\sigma_8$). As a result, this influences halo abundance, and the way halo number density responds to a change in $\sigma_8$ is quantified by $b_\phi\equiv2 \frac{\partial \ln  n}{\partial \ln \sigma_8}$.

Therefore, $b_\phi$ can be predicted if we know how $n$ (the halo mass function) changes under a rescaling of power spectrum amplitude. For a universal mass function, one finds the relation in Eq.~\eqref{eq:universality_relation} with $p=1$. If a sample of halos is preferentially selected by a secondary property, the universality assumption breaks down. Different formation histories can alter their abundance at fixed halo mass. In what follows, we parameterize this through the formation redshift $z_f$, and we should then consider a joint mass function $n=n(M, z_f)=n(M)P_{z_f}(z_f|M)$, where $P_{z_f}$ is the conditional mass function. As a result, the $b_{\phi}$ of halos of mass $M$, observed at $z_o$, that had accreted a fraction $f$ of their final mass by formation time $z_f$ is given by
\begin{equation}
 b_\phi (M,z_o,z_f,f)=2\frac{\partial \ln n(M,z_o)}{\partial \ln \sigma_8}+2\frac{\partial \ln P_{z_f}(fM,z_f|M,z_o)}{\partial \ln \sigma_8} = 2\delta_c (b_1 - p),
 \label{eq:eps}
\end{equation}

Effectively, the value of $b_\phi$ may differ from the universal value for a subpopulation of halos, because their abundance $n(M,z_f)$ does not scale with $\sigma_8$ in the same way as the global $n(M)$ does. For example, halos that assembled recently through major mergers exhibit a lower $b_{\phi}$ \cite{slosar_constraints_2008,dalal_imprints_2008}, corresponding to $p>1$. Within the extended Press-Schechter (ePS) paradigm \cite{reid_non-gaussian_2010,lacey_merger_1993,lacey_merger_1994,bosch_universal_2001} it is possible to provide a prediction for $b_{\phi}$, by writing down an analytical expression for $P_{z_f}$. \EF{This can be derived by describing halo mass accretion as a diffusion problem between redshift-dependent collapse thresholds \cite{lacey_merger_1993,lacey_merger_1994}. The solution can be written in a simplified form \cite{reid_non-gaussian_2010} by introducing the variable}

\begin{equation}\label{eq:omega_f}
    \omega_f\equiv\frac{\delta_c\left(\frac{1}{D(z_f)}-\frac{1}{D(z_o)}\right)}{\sqrt{\sigma^2(fM_o)-\sigma^2(M_o)}},
\end{equation}
that we use to parameterize assembly bias and is proportional to $z_f-z_o$ at fixed $M_o$ \cite{Fondi}. This variable depends on halo properties such as mass and formation redshift, as well as on cosmology-dependent quantities $D(z)$ and $\sigma(M)$. In terms of this variable, for $f=0.5$ \cite{lacey_merger_1993,lacey_merger_1994,reid_non-gaussian_2010} the ePS prediction for the conditional mass function reads 
\begin{equation}\label{eq:pwf_eps}
P_{\omega_f}=2\omega_f {\rm erfc}\left(\frac{\omega_f}{\sqrt{2}}\right)
\end{equation}
where ${\rm erfc}(\omega_f)=1-{\rm erf}(\omega_f)$ is the complementary error function. \EF{Early tests of the analytical expression of $P_{\omega_f}$ in Eq.~\eqref{eq:pwf_eps} against N–body simulations were performed in Refs.~\cite{reid_non-gaussian_2010,bosch_universal_2001}. These works found good agreement between Eq.~\eqref{eq:pwf_eps} and simulated halos with mass $M_o \gtrsim 2\times10^{13}\,h^{-1}M_\odot$. More recent results \cite{Fondi} based on IllustrisTNG showed that at higher redshift, the range of validity extends towards lower masses: at $z=1$, halos with $M_o \gtrsim 5\times 10^{12}\,h^{-1}M_\odot$ are accurately described by Eq.~\eqref{eq:pwf_eps}. At $z=0$, this analytical prediction breaks down, due to $P_{\omega_f}$ taking a dependence on $M_o$ and $z_o$. In this case, a fitting function with a similar form, presented in \cite{Fondi}, accurately reproduces the results. Although $\omega_f$ depends on cosmology, Eq.~\eqref{eq:pwf_eps} is effectively universal. The expression was originally derived for an EdS cosmology, yet it accurately describes results from $\Lambda$CDM simulations with different cosmologies \cite{reid_non-gaussian_2010,Fondi}, at least for the range of masses and redshift we are considering. The cosmology dependence of $p$ is fully captured in the variable $\omega_f$, which is proportional to $z_f-z_o$. Because this framework provides a physically motivated and quantitatively accurate description of the non-Gaussian halo assembly bias \cite{reid_non-gaussian_2010,Fondi}, we adopt it here.}

The step that requires care is the accurate identification of  the key characteristics of the  host halos corresponding to the tracers' sample selected observationally. This will make it possible to construct an accurate and physically motivated prior $\Pi(p)$.   
In what follows, we focus on DESI QSO sample, which is characterized by a HOD-inferred mean halo mass of $M^{\rm QSO}= 4.6\times 10^{12} \,h^{-1}M_{\odot}$ at redshift $z=1.5$ \cite{desi_hod}, falling in the regime of validity of the analytic solution.

From Eq.~\eqref{eq:eps}, using Eq.~\eqref{eq:pwf_eps} and the change of variable \eqref{eq:omega_f}, we finally obtain
\begin{equation}
 p = 1 - \frac{1}{\delta_c}\left[\frac{\omega_f \; e^{-\omega_{f}^2/2}}{\sqrt{2\pi}\,{\rm erfc}\left(\frac{\omega_f}{\sqrt{2}}\right)} - 1\right].
 \label{eq:peff}
\end{equation}
As expected, for $\omega_f=0$, corresponding to observing halos exactly at their formation $z_f=z_o$, we obtain $p=1.6$. \EF{This represents a hard upper limit for the distributions of $p$ we will show in the next Sections. }  

In the absence of a secondary dependence beyond the halo mass, the second term in Eq.~\eqref{eq:peff} vanishes and we recover universality, $p=1$.

\section{Quasars and the DESI-like sample in IllustrisTNG}
\label{sec:qso_tng}
Quasars are highly accreting, luminous active galactic nuclei (AGN), powered by accretion onto supermassive black holes. These energetic phenomena allow quasars to serve as tracers of large-scale structure, and their luminosity is directly related to the rate of matter accretion onto the black hole (BH). The accretion efficiency defines two different types of BH feedback: \textit{radio mode} and \textit{quasar mode}, which can be characterized through the Eddington ratio \cite{Sijacki07,Sijacki15,Bhowmick22}, defined as

\begin{equation}
\chi = \frac{\dot{M}_\mathrm{BH}}{\dot{M}_\mathrm{Edd}},
\label{eq:eddington_ratio}
\end{equation}
where $\dot{M}_{\rm BH}$ is the BH accretion rate and $\dot{M}_\mathrm{Edd}$ is the Eddington accretion rate, given by:
\begin{equation}
\dot{M}_\mathrm{Edd} = \frac{4 \pi G\, m_p}{\sigma_{\rm T} \,c}\frac{ M_\mathrm{BH}}{ \epsilon_r},
\label{eq:eddington_rate}
\end{equation}
with $M_\mathrm{BH}$ being the BH mass, $\epsilon_r$ the radiative efficiency (equal to $0.2$ in IllustrisTNG) and the rest are physical constants: $G$ denotes Newton constant, $m_p$ the proton mass, $c$ the speed of light and $\sigma_T$ the Thomson scattering cross-section. Ref.~\cite{Sijacki07} implemented a threshold of $\chi=0.01$ to define the quasar mode, with higher $\chi$ values corresponding to more luminous quasars.

In this Section, we make use of the TNG300-1 hydrodynamical simulation from the IllustrisTNG suite \cite{nelson_illustristng_2021}, which includes prescriptions for galaxy formation and black hole feedback, to study the properties of the dark matter halos hosting a sample of quasars mimicking the DESI QSO sample. In IllustrisTNG, the quasar mode is defined by the threshold $\chi=\min \left[\chi_0 \left(10^{-8} M^{-1}_\odot \, M_{\rm BH} \right)^\beta\, ,\, 0.1\right]$ with $\chi_0=0.002$ and $\beta=2$ \cite{Weinberger_2017}. In other words, objects with $\chi >0.1$ in IllustrisTNG are always in quasar mode.

\begin{figure}[H]
  \centering
\includegraphics[width=0.85\linewidth]{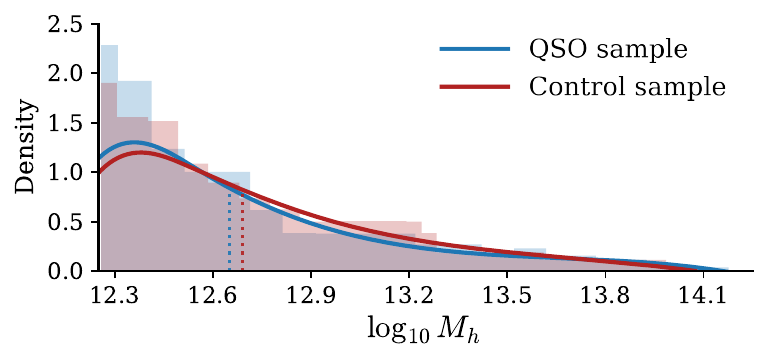}\\
  \includegraphics[width=0.85\linewidth]{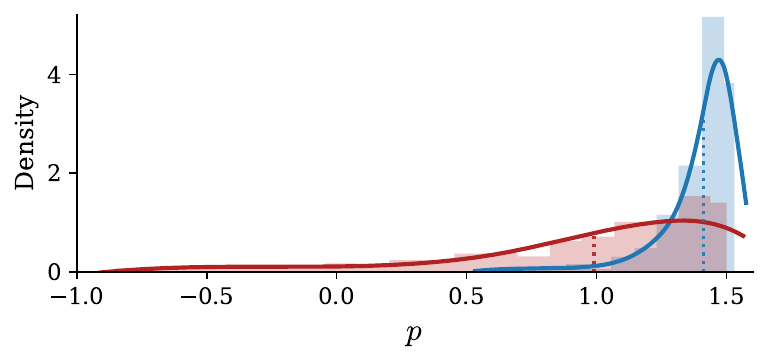}

  \caption{Distributions of host halo mass (top panel) and the $p$ parameter (bottom panel) for the QSO (in blue) and control (in red) samples. Solid curves show kernel density estimates of the normalized histograms and vertical dotted lines indicate the mean. While the two samples have (by construction) very similar halo mass distributions, the quasar hosts are skewed toward larger values of $p$, with $\bar{p}=1.41$, while $\bar{p}=0.99$ for the control sample.}
  \label{fig:distributions_qso_random}
\end{figure}

In order to select a DESI-like QSO sample, as the mean redshift of DESI QSO sample is $z_o=1.5$, we start by considering the $z_o=1.5$ halo catalog of TNG300-1. We first apply a halo mass cut $\log_{10}(M/h^{-1}M_\odot) > 12.2$ to roughly match the inferred host mass of DESI QSO sample \cite{desi_hod}. Then, to select quasar-hosting halos from this sample, we adopt a conservative approach and select the BHs with the highest $\chi$ (above $\chi=0.1$, such that they are in quasar mode). This follows from the assumption that the most efficient, and thus luminous quasars, are likely observed by surveys like DESI.
In order to match DESI QSO number density at $z_o=1.5$, we select a total of $N_q=258$ quasars from the TNG300-1 subhalo catalog. Then, we load the merger trees of the quasar-hosting halos and we measure their $z_f$ by tracing their mass accretion backwards, up to the redshift at which their mass was $0.5M_o$. The variable $\omega_f$, as defined in Eq.~\eqref{eq:omega_f}, is then computed for each halo. Finally, the parameter $p$, related to assembly bias, is determined using the ePS prediction in Eq.~\eqref{eq:peff}.

For comparison, we also select a sample of $258$ halos with a similar mass distribution of the QSO hosts, but random formation histories. This control sample is selected by minimizing the discrepancy between the two halo mass
distributions, ensuring any differences in $p$ are due to formation history, not mass.

The top panel of Fig.~\ref{fig:distributions_qso_random} shows the halo mass distributions of the two samples. By construction they are very similar: we obtain a QSO sample with mean halo mass $\log_{10} \bar{M}=12.65$ and a control sample with $\log_{10} \bar{M}=12.69$.

On the other hand, their $p$ distributions are very different, as shown in the bottom panel of Fig.~\ref{fig:distributions_qso_random}. In particular, when the halos are selected randomly (i.e., the sample is purely mass selected and samples all merger histories), the mean $p$ value is very close to 1 ($\bar{p}=0.99$), matching the prediction of the universality relation. In contrast, the $p$ distribution of the QSO sample is skewed towards large values (with $\bar{p}=1.41$), reflecting the fact that quasars predominantly form in halos with recent merger events. \EF{In the ePS model, $p$ is bounded from above by $p \leq 1.6$, so the QSO sample sits close to this upper limit and the distribution develops a long tail towards lower $p$ values.   }    .Theoretical models \cite{Hopkins_2008} have shown that mergers drive gas inflows that fuel rapid BH accretion, producing luminous quasars. Our findings are consistent with this picture, and this demonstrates that $p$ is effectively determined by the merger history rather than by halo mass.

\subsection{Robustness to selection effects}

The width of the distribution in Fig.~\ref{fig:distributions_qso_random} arises from the intrinsic scatter in the formation histories of the quasar-hosting halos within the QSO sample. 
The sample is constructed in a somewhat idealized way by imposing a sharp threshold for quasar selection, including the top $N_q$ objects ranked by $\chi$, up to some minimum $\chi_{\rm min}$. 

In reality, the DESI target selection and observational cuts may exclude some quasars or may introduce luminosity-dependent biases, where some quasars appear brighter than others, leading to imperfect selection. We test the impact of selection uncertainty on the $p$ distribution by introducing randomness in the sample selection.

Specifically, instead of enforcing a strict threshold, we introduce a probabilistic selection method. We rank the quasars by $\chi$ and define $\chi^i$ the Eddington ratio of the i-th quasar in the ranking. For this quasar, we impose a selection probability 

\begin{equation}
    P(\chi^i) = 1 - G(\chi^i),
\end{equation}

where $G$ is a Gaussian distribution centered at $\chi_{\rm min}$ with width $\sigma$. This mimics a scenario where some quasars below the threshold might scatter into the sample and vice versa. We vary $\sigma$ to achieve different fractions $r$ of objects replaced by this randomness (with larger $\sigma$ meaning more deviation from the sharp cut).

\begin{figure}[H]
  \centering
  \includegraphics[width=0.45\linewidth]{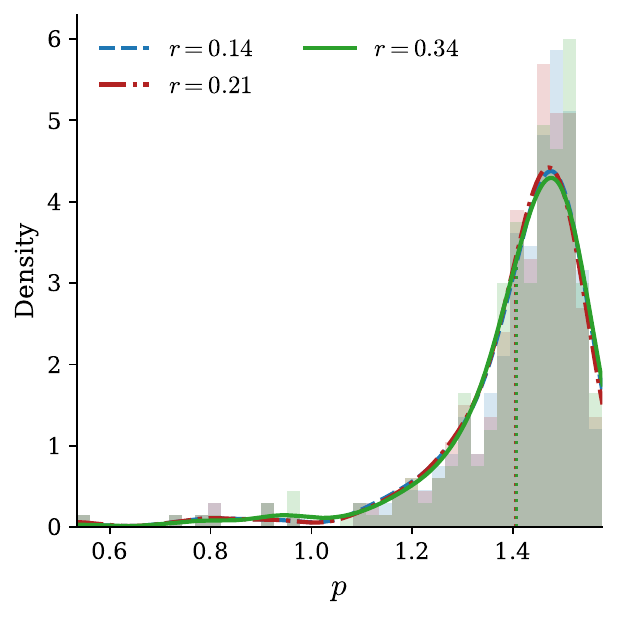}%
  \hfill
  \includegraphics[width=0.55\linewidth]{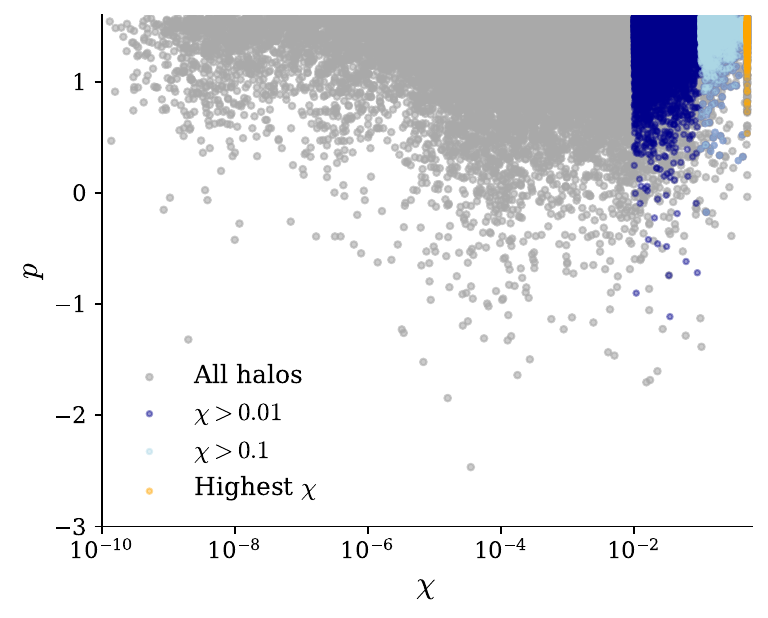}
  \caption{Left: Effect of selection noise on the $p$ distribution of the QSO sample. Faint histograms show the empirical densities and solid curves the KDE; dashed vertical lines mark the mean of each distribution. Different curves correspond to different values of $r$, the fraction of quasars replaced due to selection effects. The mean $p$ value remains stable despite introducing noise into the selection process. Right: scatter of $p$ versus Eddington ratio $\chi$. Grey points show all halos, while colored points highlight the Eddington-selected subsamples (\(\chi>0.01\), \(\chi>0.1\) -- threshold used in the literature to define the quasar mode -- and the $N_q$ highest \(\chi\)).}
  \label{fig:p-noise-and-edd}
\end{figure}

The left panel of Figure~\ref{fig:p-noise-and-edd} shows the resulting distributions of $p$ for different values of $r$. Remarkably, we find that even when 34\% of the quasars are changed with respect to the original sample, the distribution remains stable, with the same mean $\bar{p}=1.41$.

This result suggests that moderate stochasticity or biases in quasar target selection do not significantly affect the inferred $p$ for the sample.
In fact, the host halo merger history (and therefore the $p$ parameter) are similar for all halos hosting bright quasars, with a weak dependence on the specific values of the QSO Eddington accretion rate, as we show explicitly next.

In the right panel of Fig.~\ref{fig:p-noise-and-edd} we report $p$ as a function of $\chi$ for the full sample of halos at $z_o=1.5$, highlighting the selections for the thresholds $\chi=0.01$ and $ 0.1$ discussed above. As the quasar regime is approached ($\chi\!\gtrsim\!10^{-2}$), $p$ increases with $\chi$ and reaches the recent-merger expectation for the highest-$\chi$ objects. In particular, it is remarkable how the broad $\chi>0.1$ selection has the same distribution as the DESI-like QSO sample, although being $\sim\!7$ times denser. This explains why the fluctuations observed in the left panel are small: objects in quasar mode have similar formation histories, leading to a stable value of $p$, independent of small variations in selection criteria. 

\section{Validation with CAMELS Simulations: quantifying the dependence on assumptions and subgrid physics}

To assess the robustness of our results to cosmic variance and uncertainties in astrophysical modeling -- particularly those affecting black hole growth and quasar activity -- we complement our analysis with the CAMELS suite of simulations~\cite{villaescusa-navarro_camels_2021}. CAMELS (Cosmology and Astrophysics with MachinE Learning Simulations) is a set of cosmological hydrodynamical simulations designed to explore the impact of both cosmological and astrophysical parameters on structure formation, with a strong focus on subgrid physics.

In this work, we use the suite of CAMELS simulations employing  IllustrisTNG galaxy formation model. We specifically consider two sets of simulations:
\begin{itemize}
\item The \textbf{CV} set, which consists of 26 simulations run with identical cosmological and astrophysical parameters but different initial conditions. This allows us to quantify the impact of cosmic variance, treating the 26 realizations as independent sub-volumes of the Universe.
\item The \textbf{1P} set, which consists of simulations where astrophysical and cosmological parameters are varied one at a time, while the remaining ones are held fixed. Each parameter is sampled at 4 values different from the fiducial IllustrisTNG counterpart. We focus on variations of 9 parameters controlling AGN feedback, BH growth and QSO activity, corresponding to a total of 36 realizations. The parameters, together with their range of variation and description, are listed in Table~~\ref{tab:agn_params}. This set allows us to explore how sensitive our QSO $p$ distribution is to the assumed parameterization in IllustrisTNG, effectively marginalizing over the implementation of AGN and QSO physics. 
\end{itemize}
\EF{Although the full CAMELS-1P suite also includes realizations where cosmological parameters are varied, we do not use them here. Our goal is to quantify the impact of subgrid astrophysical modelling on the $p$ prior at fixed cosmology, consistently with the DESI DR1 analysis we update, which assumes a fiducial cosmology for the linear power spectrum.  The robustness to assumptions about the underlying cosmology has been demonstrated in e.g., \cite{Fondi}. }.

\begin{table}[H]
\centering
\begin{tabular}{lccc}
\toprule
\textbf{Parameter} & \textbf{Fiducial} & \textbf{Range} & \textbf{Description} \\
\midrule
$A_{\rm AGN1}$ & 1 & [0.25, 4] & Low-accretion feedback energy \\
$A_{\rm AGN2}$ & 20 & [10, 40] & Feedback burstiness, $f_{re}$ in \cite{Weinberger_2017} \\
$M_{\rm BH, 0}$ & 8 & [2.5, 25] & BH seed mass (units of $10^5 M_{\odot}$) \\
$A_{\dot{M}_{\rm BH}}$ & 1 & [0.25, 4.0] & $\dot{M}_{\rm BH}$ multiplier \\
$A_{\dot{M}_{\rm Edd}}$ & 1 & [0.1, 10.0] & $\dot{M}_{\rm Edd}$ multiplier \\
$\epsilon_{f, \rm high}$ & 0.1 & [0.025, 0.4] & High-accretion feedback energy, Eq.~7 in \cite{Weinberger_2017} \\
$\epsilon_{r}$ & 0.2 & [0.05, 0.8] & Radiative efficiency \\
$\chi_0$ & 0.002 & [$0.000063$, 0.063] & $\chi$ threshold amplitude, Eq.~5 in \cite{Weinberger_2017} \\
$\beta$ & 2 & [0.0, 4.0] & $\chi$ threshold index, Eq.~5 in \cite{Weinberger_2017} \\
\bottomrule
\end{tabular}
\caption{Astrophysical parameters varied in the CAMELS-1P set. Each parameter is varied individually while others are held fixed. In the second column, we report the fiducial IllustrisTNG value, while the range of variation of the parameters in CAMELS is reported in the third one. A brief description is displayed in the last column.}
\label{tab:agn_params}
\end{table}

Each CAMELS volume (with size $L=50\,h^{-1}$Mpc) is $\sim 68$ times smaller than TNG300-1. As a consequence, to match DESI number density at $z\sim1.5$, \EF{we select the top $4$ quasars per box after rank-ordering them by their $\chi$. This ensures that we always select the most highly accreting quasars in a way that is robust to changes in AGN and QSO subgrid parameters, while matching the DESI QSO abundance.}.  By combining the different realizations, we then obtain a total of $N_{\rm CV}=104$ and $N_{\rm 1P}=144$ objects for the CV and 1P sets, respectively. 

\begin{figure}[H]
\begin{center}
\includegraphics[width=0.8\linewidth]{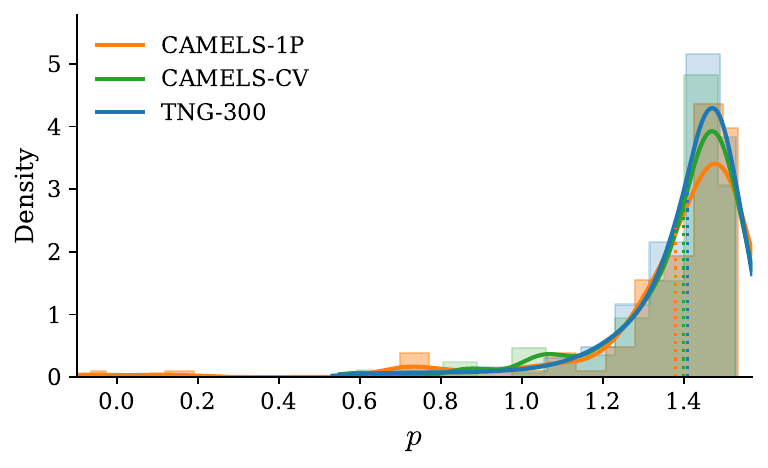}
\end{center}
\caption{Empirical distributions of the parameter $p$ for DESI-like QSO selections in simulations. Filled histograms show the stacked samples; smooth curves are KDE estimates. Blue dashed: TNG-300; green: CAMELS-CV (cosmic-variance set); orange: CAMELS-1P (astrophysics-variation set). The vertical dotted lines mark the mean of the distributions.}
\label{fig:p_distribution_cv}
\end{figure}

The stacked distribution of $p$ values across CAMELS-CV and CAMELS-1P sets is shown in Fig.~\ref{fig:p_distribution_cv}, where we also report the IllustrisTNG distribution, for comparison.
The CAMELS-CV distribution closely follows the TNG-300 result, showing a small impact of cosmic variance.

The 1P distribution is instead broadened by the presence of objects with slightly different merger histories than in the fiducial case. This is a result of the different AGN feedback parameterizations, as well as the QSO activities, which affect their selection. However, this effect is relatively small and the overall shape of the distribution is preserved, with a small shift of the mean from $\bar{p}=1.41$ to $\bar{p}=1.38$.

The results confirm that the enhancement in $p$ due to recent merger history is not a byproduct of a specific choice of subgrid physics. Even after marginalizing over astrophysical parameters, the $p$ distribution remains peaked at $p > 1$, in line with expectations for quasar samples dominated by recently merged halos. This reinforces the robustness of our prior on $p$ and supports its application in cosmological analyses based on DESI QSO clustering.

\section{Updated DESI DR1 constraints}
\label{sec:fnl_constraints}
In this Section we use our simulation-based prior $\Pi(p)$ -- the probability distribution of $p$ for the relevant tracers -- to update DESI DR1 $f_{\rm NL}$ constraints from the QSO sample. Constraints on $f_{\rm NL}$ are obtained from power spectrum measurements using the model
\begin{equation}
P(k,\mu)
= \frac{\left[
    b_1(z_{\rm eff})
    + \frac{b_\phi(z_{\rm eff})}{T_{\Phi\to\delta}(k,z_{\rm eff})}\, f_{\rm NL}
    + f(z_{\rm eff})\,\mu^2
  \right]^2}{
  \left[1+\tfrac{1}{2}\,(k\mu\Sigma_s)^2\right]^{2}}
  P_{\rm lin}(k,z_{\rm eff})
  + s_{n,0}\,,
\end{equation}
which includes the Kaiser effect and a damping factor, determined by $\Sigma_s$, to model redshift space distortions. $f$ is the linear growth rate and $s_{n,0}$ is a potential residual shot noise. The redshift-dependent quantities are evaluated at the effective redshift $z_{\rm eff}$, which depends on the \textit{weighting scheme} adopted for the power spectrum estimator. We refer the reader to Ref.~\cite{desi_fnl2024} for more details.
For the scope of this work, it is sufficient to mention that for the optimal quadratic estimator (OQE) weights, quasars are weighted by
\begin{equation}
\label{eq:weights}
 \tilde{w}(z)=b_1(z)-p_w \,,  
\end{equation}
where the linear bias $b_1(z)$ is given by a fitting function calibrated in Ref.~\cite{desi_fnl2024} and $p_w$ is the parameter used for the OQE weights. DR1 analysis used two choices $p_w=1$ and $p_w=1.6$, which yield slightly different effective redshifts for the QSO sample, $z_{\rm eff}=1.93$ and $z_{\rm eff}=2.08$, respectively. Correspondingly, we repeat the procedure outlined in the previous sections to build the prior on $p$ from the snapshots closest to these redshifts, both for IllustrisTNG and CAMELS. In Appendix \ref{app:redshift_dependence}, we report the redshift dependence of the $p$ distribution in IllustrisTNG, which is found to be very weak. 

\begin{figure}[H]
  \centering
  \includegraphics[width=0.53\linewidth]{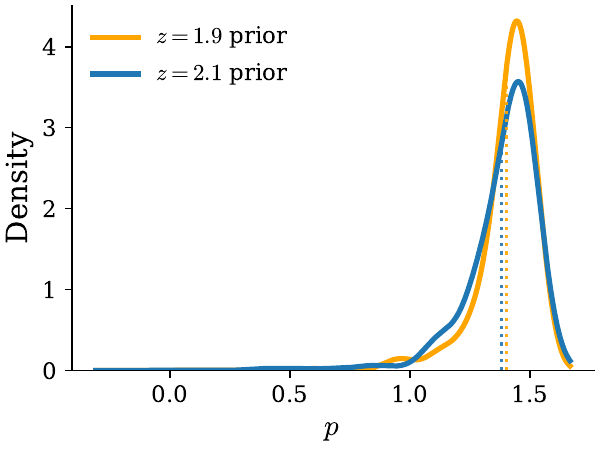}%
  \raisebox{3mm}{
  \includegraphics[width=0.4\linewidth]{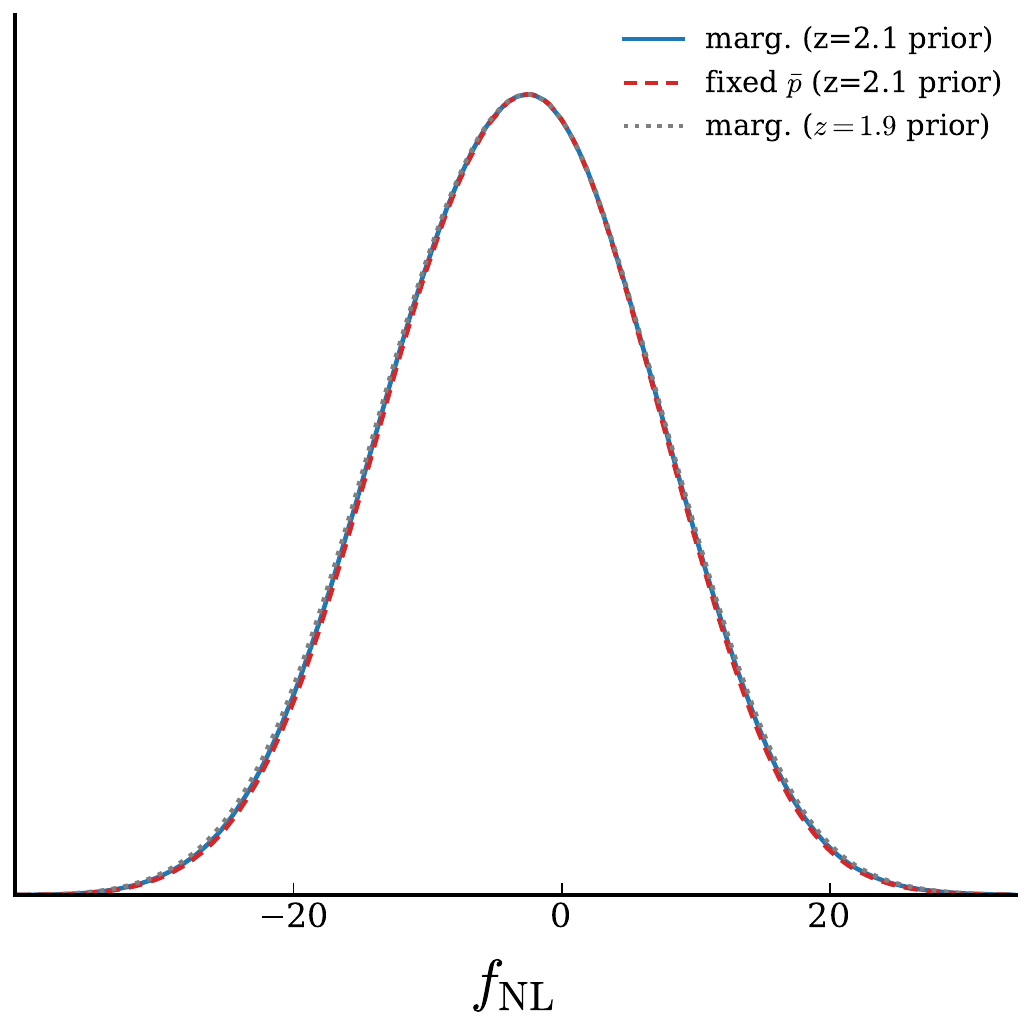}
  }
  \caption{Left: Prior distributions of the assembly–bias parameter $p$ built from the DESI–like QSO selections in IllustrisTNG+CAMELS (combined) at the two OQE effective redshifts. The curves show KDE–smoothed distributions for $z_{\rm eff}\simeq1.9$ (orange) and $z_{\rm eff}\simeq2.1$ (blue); vertical dotted lines mark the corresponding means. Right: One–dimensional posterior for $f_{\rm NL}$ from DESI DR1 QSOs $[b_1,b_{\phi}f_{\rm NL}]$ chains with $p_w=1.6$ OQE weights. Three choices are shown: marginalizing over the $z\simeq2.1$ prior on $p$ (solid blue), fixing $p$ to the mean of that prior (dashed red), and marginalizing over the $z\simeq1.9$ prior (dotted gray).}
  \label{fig:fnl_marginalization}
\end{figure}

As a first test, we start from the DR1 MCMC chains obtained from sampling the joint posterior of $\left[b_1,b_{\phi}f_{\rm NL}\right]$. We propagate our prior on $p$ by marginalization: for each chain sample $(b_1,b_{\phi}f_{\rm NL})$ and for each $p_j$ sampled from the empirical prior $\Pi(p_j)$ we compute $b_\phi(p_j)=2\delta_c(b_1-p_j)$ and obtain $f_{{\rm NL},j}=(b_{\phi}f_{\rm NL})/b_\phi(p_j)$. Stacking all $j$ yields the marginalized posterior for $f_{\rm NL}$.
The left panel of Fig.~\ref{fig:fnl_marginalization} shows the prior $\Pi(p)$ at the two effective redshifts. Each $\Pi$ is built by combining the DESI-like QSO samples from IllustrisTNG and CAMELS at that $z_{\rm eff}$ (i.e., we simply pool the two $p$ catalogs). For the marginalization we use directly the $N$ raw $p$ values from this combined sample, so the result is independent of histogram binning or KDE bandwidth; this is equivalent to drawing $N$ samples from an analytic prior but avoids any kernel/parametric choices. This combination naturally captures the spread induced by astrophysical uncertainty in CAMELS.

On the right panel, we display the $f_{\rm NL}$ posterior from the $[b_1,b_{\phi}f_{\rm NL}]$ chains with $p_w=1.6$ OQE weights ($z_{\rm eff}\simeq2.1$), for three different assumptions. Specifically, we verify that marginalizing over the full prior or simply fixing $p$ to the prior mean yields negligible differences. Moreover, we also explore the effect of marginalizing over the $z_{\rm eff}=1.9$ prior, showing that the result is insensitive to the small redshift dependence of the prior. \EF{These statistically indistinguishable differences are a consequence of the current level of precision, related to other sources of uncertainties. Despite the skewness of $\Pi(p)$, replacing the full prior by a fixed value $p=\langle p\rangle$ leads to indistinguishable constraints on $f_{\rm NL}$, and choosing the  the median instead of the mean would also have negligible impact on DR1 constraints.} 

\begin{figure}[H]
\begin{center}
\includegraphics[width=0.9\linewidth]{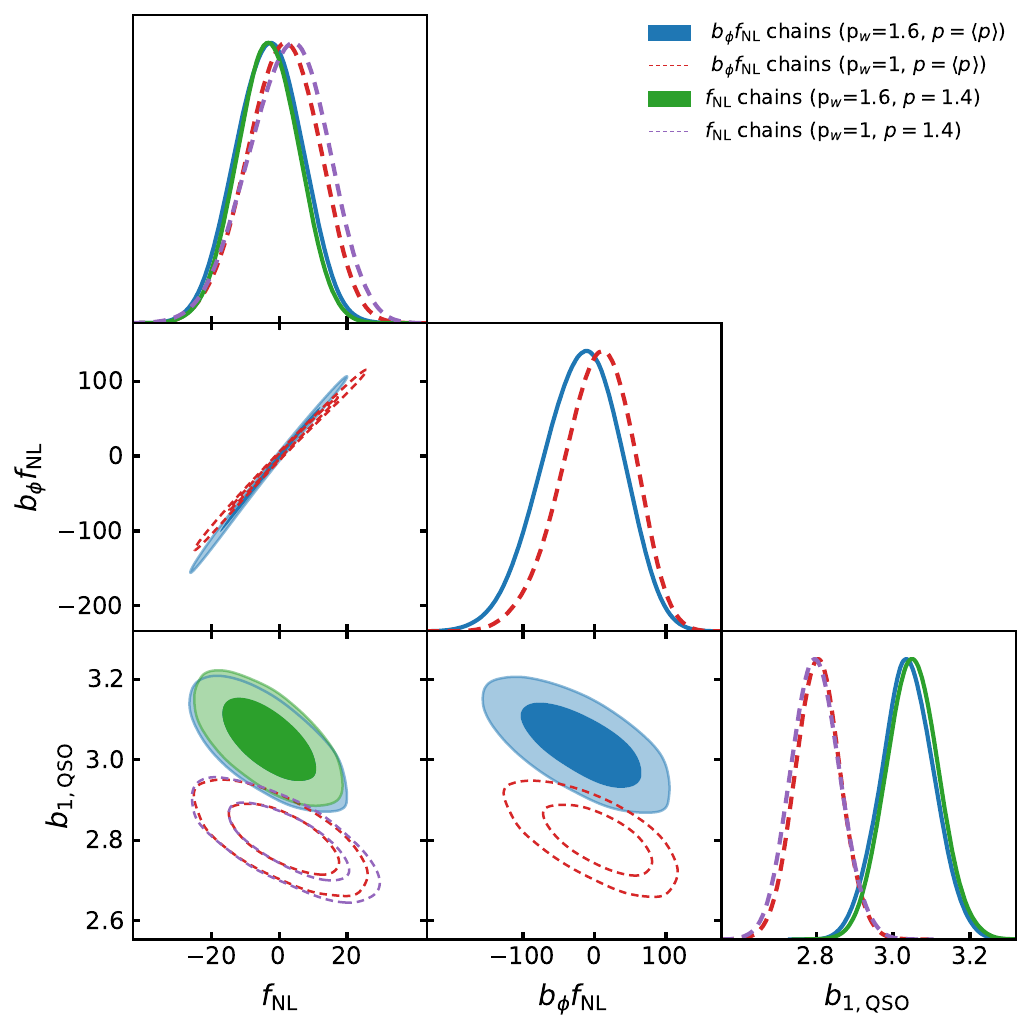}
\end{center}
\caption{
Corner plot comparing different chains and weight choices for DESI DR1 QSOs $f_{\rm NL}$ inference. Shaded/solid contours correspond to analyses run with $p_w=1.6$; while dashed contours use $p_w=1$. Blue (solid) and red (dashed) curves show chains in which we sample $[b_\phi f_{\rm NL},\,b_1]$ and obtain $f_{\rm NL}$ constraints by fixing $p$ to the prior mean within these samples. Green (solid) and purple (dashed) show chains in which $p=1.4$ is fixed during inference and $[f_{\rm NL},\,b_1]$ are sampled.}
\label{fig:triangle_plot}
\end{figure}

Finally, we run the chains by fixing $p=1.4$ during inference and sample $\left[b_1,f_{\rm NL}\right]$. The constraints are reported in Fig.~\ref{fig:triangle_plot}, in which we also show the contours associated with $[b_1,b_{\phi}f_{\rm NL}]$ chains. 

Figure~\ref{fig:triangle_plot} highlights the key parameter degeneracies and the impact of the OQE weighting choice. As expected, the $(b_\phi f_{\rm NL},\,f_{\rm NL})$ panel exhibits a tight correlation since $f_{\rm NL}$ and $b_\phi$ are largely degenerate before imposing a prior on $p$. 
The $b_1$ posteriors are shifted between weightings: $b_1$ is systematically larger for $p_w=1.6$ than for $p_w=1$, consistent with the higher effective redshift of the $p_w=1.6$ sample and the known increase of bias with redshift. The 1D $f_{\rm NL}$ posteriors show a dependence on the weighting ($p_w = 1$ vs.\ $1.6$), reflecting the modest change in $z_{\rm eff}$.

\EF{ In principle the slight inconsistency between the choices of $p_w$ and $p$ can bias the recovered $f_{\rm NL}$. This is not an issue at the current level of precision, as the bias is at the 1/10 $\sigma$ level. 
Re-running the full power-spectrum pipeline with $p_w=1.4$ would require recomputing the weighted power spectrum, window matrix and mock catalogues, which is computationally expensive. Instead, we use the existing DR1 measurements and interpret the shift between the two posteriors with $p_w=1$ and $p_w=1.6$ as an empirical upper bound on the possible bias. For future DESI data releases, our results motivate adopting $p_w=1.4$ directly in the measurement and mock pipelines, thereby ensuring a fully self-consistent treatment of $p$. }
 
 Finally, comparing the blue/red solid contours (post-processing the $[b_1, b_\phi f_{\rm NL}]$ chains by conditioning on $p=\langle p\rangle$) to the green/purple contours
\EF{ (re-running the sampler and fixing $p=1.4$ during inference while sampling $[b_1, f_{\rm NL}]$ directly) }

shows that the two approaches are statistically indistinguishable.

\section{Conclusions}
Large-scale structure constraints on local PNG rely on the scale-dependent bias induced in two-point clustering, which is sensitive to the product $f_{\rm NL}\,b_\phi$. Crucially $b_\phi$ depends on the sample linear bias $b_1$ which can be estimated from the data themselves, and the parameter $p$ which is usually not known and notoriously  hard to measure. The standard universality relation, Eq.~\eqref{eq:universality_relation} with $p=1$, is only valid for fair, halo mass-selected samples, while realistic tracer selections are generally affected by assembly bias, having $p\neq1$. In DESI DR1 this uncertainty was handled by fixing $p$ to either $1$ or $1.6$, yielding two separate $f_{\rm NL}$ constraints.

In this work, we have built a physically motivated, simulation-based prior for the assembly bias parameter $p$ of DESI quasars and used it to obtain a single, robust $f_{\rm NL}$ constraint from DESI DR1 QSO sample.

From a DESI–like selection in IllustrisTNG TNG300-1, quasar hosts are preferentially recent mergers, yielding a $p$ distribution skewed to values $>1$ with mean $\bar{p}\simeq 1.4$ at $z\simeq 1.5$. This result is robust to selection noise (we find no shift in $\bar p$ even when $\sim 34\%$ of objects are replaced) and is consistent with previous assumptions on the merger history of quasar-hosting halos. We have validated our prior with CAMELS simulations, which allow us to assess the dependence on the specific implementation of subgrid physics in IllustrisTNG. The results show a negligible impact from cosmic variance and only a mild broadening when varying AGN/QSO subgrid parameters, with the mean shifting by 1.4\% to $\bar p\simeq 1.38$. The redshift evolution of the prior is weak  enough over the range relevant for DR1 to make it negligible at the current level of precision afforded  by the data. Together, these tests support the use of an empirical prior $\Pi(p)$ for DESI QSOs, built by combining IllustrisTNG and CAMELS distributions, which we report in Section \ref{sec:fnl_constraints}.

We propagate this prior through $b_\phi=2\delta_c\,(b_1-p)$ in the DESI DR1 analysis of \cite{desi_fnl2024}. Because the OQE weights used in \cite{desi_fnl2024} depend on the assumed value of the $p$ parameter, $p_w$ defined in Eq.\eqref{eq:weights}, we rebuild $\Pi(p)$ at the two effective redshifts corresponding to the DR1 choices ($p_w=1$ and $p_w=1.6$). Post-processing the DR1 $[b_1, b_\phi f_{\rm NL}]$ chains by marginalizing over $\Pi(p)$ yields $f_{\rm NL}$ posteriors that are indistinguishable from those obtained by re-running the DESI DR1 sampler, fixing $p$ to the prior mean and sampling $[b_1, f_{\rm NL}]$ directly. 

\begin{table}[t]
\centering
\begin{tabular}{lcc}
\hline
$p_w$ & $p$ & $f_{\rm NL}$ \\
\hline
\hline
$1.6$ & $1.4$ & $-3.3 \pm 9.2$ \\
$1.6$ & $1.6$ & $-2^{+11}_{-10}$ \cite{desi_fnl2024} \\
$1$ & $1.4$ & $2.4 \pm 11.7$ \\
$1$ & $1$ & $3.5^{+10.7}_{-7.4}$ \cite{desi_fnl2024} \\
\hline
\end{tabular}
\caption{$f_{\rm NL}$ constraints obtained in this work and in the DESI DR1 analysis of Ref.~\cite{desi_fnl2024}. For each value of the OQE weighting parameter $p_w$, we report our result using the prior-informed value $p=1.4$, and the corresponding DR1 result where $p$ was set equal to $p_w$.
}
\label{tab:oqe_comparison}
\end{table}

\EF{In Table~\ref{tab:oqe_comparison} we compare our updated constraints with the original DESI DR1 results \cite{desi_fnl2024}. For our fiducial choice of OQE weights, $p_w=1.6$, adopting the prior-informed value $p=1.4$ yields slightly tighter constraints, due to a larger value of $b_{\phi}$, and vice-versa for $p_w=1$. This comparison shows that updating the value of $p$ using our simulation-based prior has only a modest impact at the current DR1 precision, while providing a more physically motivated description of the QSO sample that will become increasingly relevant for future DESI data releases.}

\section*{Data Availability}
The data used in this work are publicly available as part of DESI Data Release 1 (see \url{https://data.desi.lbl.gov/doc/releases/}). The data points corresponding to the figures are available on Zenodo at \href{https://zenodo.org/records/18469267}{10.5281/zenodo.18469267}  in compliance with the DESI data management plan.

\section*{Acknowledgments}
EF thanks Francisco Villaescusa-Navarro for providing access to the CAMELS simulations. EF acknowledges the support from ``la Caixa” Foundation (ID 100010434, code LCF/BQ/DI21/ 11860061).
Funding for this work was partially provided by project PID2022-141125NB-I00MCIN/ AEI/10.13039 /501100011033  and  “Center of Excellence Maria de Maeztu" award to the ICCUB CEX2024-001451-M funded by MICIU/AEI/10.13039/501100011033.

This material is based upon work supported by the U.S.\ Department of Energy (DOE), Office of Science, Office of High-Energy Physics, under Contract No.\ DE–AC02–05CH11231, and by the National Energy Research Scientific Computing Center, a DOE Office of Science User Facility under the same contract. Additional support for DESI was provided by the U.S. National Science Foundation (NSF), Division of Astronomical Sciences under Contract No.\ AST-0950945 to the NSF National Optical-Infrared Astronomy Research Laboratory; the Science and Technology Facilities Council of the United Kingdom; the Gordon and Betty Moore Foundation; the Heising-Simons Foundation; the French Alternative Energies and Atomic Energy Commission (CEA); the National Council of Humanities, Science and Technology of Mexico (CONAHCYT); the Ministry of Science and Innovation of Spain (MICINN), and by the DESI Member Institutions: \url{https://www.desi. lbl.gov/collaborating-institutions}. 

The DESI Legacy Imaging Surveys consist of three individual and complementary projects: the Dark Energy Camera Legacy Survey (DECaLS), the Beijing-Arizona Sky Survey (BASS), and the Mayall z-band Legacy Survey (MzLS). DECaLS, BASS and MzLS together include data obtained, respectively, at the Blanco telescope, Cerro Tololo Inter-American Observatory, NSF NOIRLab; the Bok telescope, Steward Observatory, University of Arizona; and the Mayall telescope, Kitt Peak National Observatory, NOIRLab. NOIRLab is operated by the Association of Universities for Research in Astronomy (AURA) under a cooperative agreement with the National Science Foundation. Pipeline processing and analyses of the data were supported by NOIRLab and the Lawrence Berkeley National Laboratory. Legacy Surveys also uses data products from the Near-Earth Object Wide-field Infrared Survey Explorer (NEOWISE), a project of the Jet Propulsion Laboratory/California Institute of Technology, funded by the National Aeronautics and Space Administration. Legacy Surveys was supported by: the Director, Office of Science, Office of High Energy Physics of the U.S. Department of Energy; the National Energy Research Scientific Computing Center, a DOE Office of Science User Facility; the U.S. National Science Foundation, Division of Astronomical Sciences; the National Astronomical Observatories of China, the Chinese Academy of Sciences and the Chinese National Natural Science Foundation. LBNL is managed by the Regents of the University of California under contract to the U.S. Department of Energy. The complete acknowledgments can be found at \url{https://www.legacysurvey.org/}.

Any opinions, findings, and conclusions or recommendations expressed in this material are those of the author(s) and do not necessarily reflect the views of the U.S.\ National Science Foundation, the U.S.\ Department of Energy, or any of the listed funding agencies.

The authors are honored to be permitted to conduct scientific research on I’oligam Du’ag (Kitt Peak), a mountain with particular significance to the Tohono O’odham Nation.

\appendix
\section{Redshift dependence}
\label{app:redshift_dependence}

To assess whether the parameter \( p \) exhibits any redshift dependence, we repeat the same procedure described in Section \ref{sec:qso_tng} at different redshifts. Specifically, we select quasar-hosting halos at various redshifts and determine their \( p \) values using Eq.~\eqref{eq:peff}. 

In order to match the DESI number density $n^{\rm QSO}(z)$ at each redshift, we use the tabulated values from DESI Year 1 (Y1) data and select $N_q=n^{\rm QSO}(z) \; \cdot V$ quasars, where $V$ is the volume of the simulation box.

\begin{figure}[H]
\begin{center}
	\includegraphics[width=0.9\linewidth]{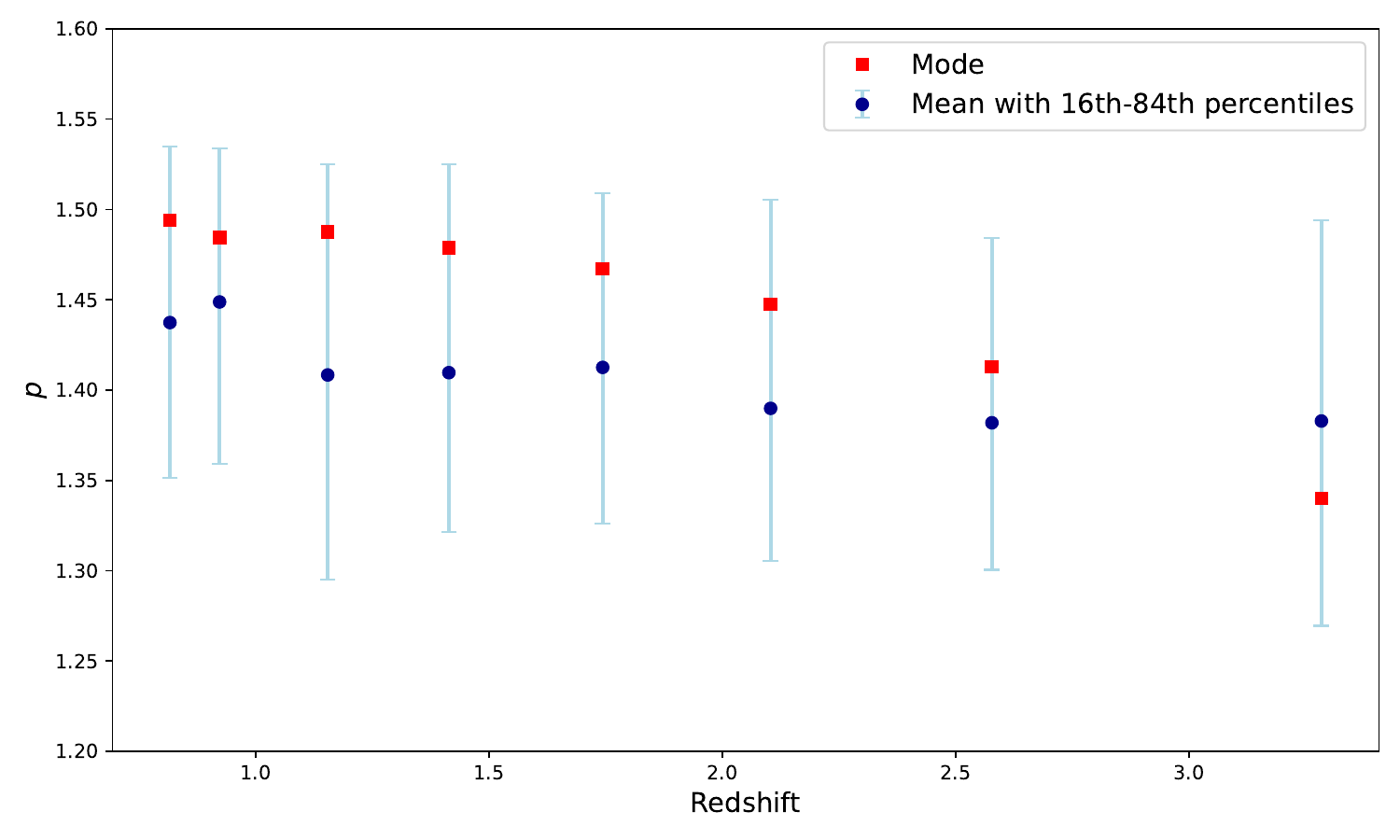}
\end{center}
\caption{Evolution of \( p \) as a function of redshift. At each redshift, the red point represents the mode of the distribution, while the light blue bar  denotes the 16th-84th percentiles range around the mean value, shown in dark blue.}
\label{fig:p_redshift}
\end{figure}

As previously observed in Fig.~\ref{fig:distributions_qso_random}, the distribution of \( p \) is non-Gaussian. To summarize the variation of \( p \) across redshifts, in Figure~\ref{fig:p_redshift} we report the mean and the 16th-84th percentile range of the distributions, as well as the mode.

From Figure~\ref{fig:p_redshift}, we observe a weak dependence of \( p \) on redshift, with the mean of the distribution remaining stable. The trend observed in Figure~\ref{fig:p_redshift} appears coherent with expectations, since the merger history of quasars is more intense at redshifts below \( z\sim2 \), leading to an expected increase in \( p \) towards $1.6$ as \( z \) decreases. Conversely, at high redshifts, where quasars typically exhibit weaker merger activity, one would naively expect \( p \) to decrease.

The observed redshift dependence is very weak, especially compared to the width of the distribution, and in many practical applications it can be neglected.

\section{Affiliations}
\label{app:affiliations}

\textsuperscript{1}{Institut de Ci\`encies del Cosmos, Universitat de Barcelona (ICCUB), c. Mart\'i i Franqu\`es, 1, 08028 Barcelona, Spain.}
\\ \textsuperscript{2}{Instituci\'{o} Catalana de Recerca i Estudis Avan\c{c}ats, Passeig de Llu\'{\i}s Companys, 23, 08010 Barcelona, Spain}
\\ \textsuperscript{3}{Lawrence Berkeley National Laboratory, 1 Cyclotron Road, Berkeley, CA 94720, USA}
\\ \textsuperscript{4}{Department of Physics, Boston University, 590 Commonwealth Avenue, Boston, MA 02215 USA}
\\ \textsuperscript{5}{Department of Physics \& Astronomy, University of Rochester, 206 Bausch and Lomb Hall, P.O. Box 270171, Rochester, NY 14627-0171, USA}
\\ \textsuperscript{6}{Dipartimento di Fisica ``Aldo Pontremoli'', Universit\`a degli Studi di Milano, Via Celoria 16, I-20133 Milano, Italy}
\\ \textsuperscript{7}{INAF-Osservatorio Astronomico di Brera, Via Brera 28, 20122 Milano, Italy}
\\ \textsuperscript{8}{Department of Physics \& Astronomy, University College London, Gower Street, London, WC1E 6BT, UK}
\\ \textsuperscript{9}{Instituto de F\'{\i}sica, Universidad Nacional Aut\'{o}noma de M\'{e}xico,  Circuito de la Investigaci\'{o}n Cient\'{\i}fica, Ciudad Universitaria, Cd. de M\'{e}xico  C.~P.~04510,  M\'{e}xico}
\\ \textsuperscript{10}{University of California, Berkeley, 110 Sproul Hall \#5800 Berkeley, CA 94720, USA}
\\ \textsuperscript{11}{Departamento de F\'isica, Universidad de los Andes, Cra. 1 No. 18A-10, Edificio Ip, CP 111711, Bogot\'a, Colombia}
\\ \textsuperscript{12}{Observatorio Astron\'omico, Universidad de los Andes, Cra. 1 No. 18A-10, Edificio H, CP 111711 Bogot\'a, Colombia}
\\ \textsuperscript{13}{Institut d'Estudis Espacials de Catalunya (IEEC), c/ Esteve Terradas 1, Edifici RDIT, Campus PMT-UPC, 08860 Castelldefels, Spain}
\\ \textsuperscript{14}{Institute of Cosmology and Gravitation, University of Portsmouth, Dennis Sciama Building, Portsmouth, PO1 3FX, UK}
\\ \textsuperscript{15}{Institute of Space Sciences, ICE-CSIC, Campus UAB, Carrer de Can Magrans s/n, 08913 Bellaterra, Barcelona, Spain}
\\ \textsuperscript{16}{University of Virginia, Department of Astronomy, Charlottesville, VA 22904, USA}
\\ \textsuperscript{17}{Fermi National Accelerator Laboratory, PO Box 500, Batavia, IL 60510, USA}
\\ \textsuperscript{18}{Institut d'Astrophysique de Paris. 98 bis boulevard Arago. 75014 Paris, France}
\\ \textsuperscript{19}{IRFU, CEA, Universit\'{e} Paris-Saclay, F-91191 Gif-sur-Yvette, France}
\\ \textsuperscript{20}{Department of Physics, University of Michigan, 450 Church Street, Ann Arbor, MI 48109, USA}
\\ \textsuperscript{21}{University of Michigan, 500 S. State Street, Ann Arbor, MI 48109, USA}
\\ \textsuperscript{22}{Department of Physics, The University of Texas at Dallas, 800 W. Campbell Rd., Richardson, TX 75080, USA}
\\ \textsuperscript{23}{NSF NOIRLab, 950 N. Cherry Ave., Tucson, AZ 85719, USA}
\\ \textsuperscript{24}{The Ohio State University, Columbus, 43210 OH, USA}
\\ \textsuperscript{25}{Sorbonne Universit\'{e}, CNRS/IN2P3, Laboratoire de Physique Nucl\'{e}aire et de Hautes Energies (LPNHE), FR-75005 Paris, France}
\\ \textsuperscript{26}{Departament de F\'{i}sica, Serra H\'{u}nter, Universitat Aut\`{o}noma de Barcelona, 08193 Bellaterra (Barcelona), Spain}
\\ \textsuperscript{27}{Institut de F\'{i}sica d'Altes Energies (IFAE), The Barcelona Institute of Science and Technology, Edifici Cn, Campus UAB, 08193, Bellaterra (Barcelona), Spain}
\\ \textsuperscript{28}{Center for Cosmology and AstroParticle Physics, The Ohio State University, 191 West Woodruff Avenue, Columbus, OH 43210, USA}
\\ \textsuperscript{29}{Department of Astronomy, The Ohio State University, 4055 McPherson Laboratory, 140 W 18th Avenue, Columbus, OH 43210, USA}
\\ \textsuperscript{30}{Department of Physics and Astronomy, University of Waterloo, 200 University Ave W, Waterloo, ON N2L 3G1, Canada}
\\ \textsuperscript{31}{Perimeter Institute for Theoretical Physics, 31 Caroline St. North, Waterloo, ON N2L 2Y5, Canada}
\\ \textsuperscript{32}{Waterloo Centre for Astrophysics, University of Waterloo, 200 University Ave W, Waterloo, ON N2L 3G1, Canada}
\\ \textsuperscript{33}{Instituto de Astrof\'{i}sica de Andaluc\'{\i}a (CSIC), Glorieta de la Astronom\'{\i}a, s/n, E-18008 Granada, Spain}
\\ \textsuperscript{34}{Departament de F\'isica, EEBE, Universitat Polit\`ecnica de Catalunya, c/Eduard Maristany 10, 08930 Barcelona, Spain}
\\ \textsuperscript{35}{Department of Physics and Astronomy, Sejong University, 209 Neungdong-ro, Gwangjin-gu, Seoul 05006, Republic of Korea}
\\ \textsuperscript{36}{Abastumani Astrophysical Observatory, Tbilisi, GE-0179, Georgia}
\\ \textsuperscript{37}{Department of Physics, Kansas State University, 116 Cardwell Hall, Manhattan, KS 66506, USA}
\\ \textsuperscript{38}{Faculty of Natural Sciences and Medicine, Ilia State University, 0194 Tbilisi, Georgia}
\\ \textsuperscript{39}{CIEMAT, Avenida Complutense 40, E-28040 Madrid, Spain}
\\ \textsuperscript{40}{National Astronomical Observatories, Chinese Academy of Sciences, A20 Datun Road, Chaoyang District, Beijing, 100101, P.~R.~China}

\providecommand{\href}[2]{#2}\begingroup\raggedright\endgroup
\end{document}